\documentclass[aps,twocolumn,
footinbib]{revtex4-1} 
\usepackage{amsmath,amssymb,bm}
\usepackage{graphicx}
\usepackage{epstopdf}
\usepackage{latexsym}
\usepackage[normalem]{ulem} 
\usepackage[caption=false]{subfig}
\usepackage[usenames,dvipsnames]{color}
\usepackage{hyperref}
\usepackage{natbib}
\usepackage{bbold}

\usepackage[utf8]{inputenc} 

\newcommand{\ket}[1]{\left|#1\right\rangle}

\newcommand{\ind}[1]{_\text{#1}}

\begin{document}

\title{Classical and quantum chaos in a three-mode bosonic system}

\date{\today}

\author{Michael Rautenberg}
\author{Martin G\"{a}rttner}
\affiliation{Kirchhoff-Institut f\"{u}r Physik, Universit\"{a}t Heidelberg, Im Neuenheimer Feld 227, 69120 Heidelberg, Germany
}

\begin{abstract}
We study the dynamics of a three-mode bosonic system with mode-changing interactions. 
For large mode occupations the short-time dynamics is well described by classical mean-field equations allowing us to study chaotic dynamics in the classical system and its signatures in the corresponding quantum dynamics. By introducing a symmetry-breaking term we tune the classical dynamics from integrable to strongly chaotic which we demonstrate by calculating Poincar\'e sections and Lyapunov exponents. The corresponding quantum system features level statistics that change from Poissonian in the integrable to Wigner-Dyson in the chaotic case. We investigate the behavior of out-of-time-ordered correlators (OTOCs), specifically the squared commutator, for initial states located in regular and chaotic regions of the classical mixed phase space and find marked differences between the two cases.
The short-time behavior is well captured by semi-classical truncated Wigner simulations directly relating these features to properties of the underlying classical mean field dynamics.
We discuss a possible experimental realization of this model system in a Bose-Einstein condensate of rubidium atoms which allows reversing the sign of the Hamiltonian required for measuring OTOCs experimentally.
\end{abstract}

\maketitle  

\section{Introduction}
\label{sec:intro}

Statistical mechanics is based on the assumption that in thermal equilibrium all energetically accessible microstates of a system are populated with equal probability. This assumption allows to derive a comprehensive description of equilibrium phenomena. In contrast, the question how equilibrium is reached if the system is prepared in a state far from equilibrium turns out to be rather difficult to answer in general. For classical systems the answer is closely connected to the notion of classical chaos. If trajectories in phase space are not closed and a given initial state eventually explores the whole accessible phase space volume, then the system thermalizes in the sense that the long-time average of an observable is equal to the microcanonical average \cite{penrose1979}. This notion of chaos also entails the exponential sensitivity to the initial condition, i.e. that initially close trajectories will deviate exponentially from each other at later times, which can be quantified by the Lyapunov exponent \cite{Eckmann2004}.

For quantum systems the relation between the microscopic dynamics and thermodynamic ensembles is more subtle since the notion of trajectories in phase space is in general absent due to Heisenberg's uncertainty principle. Moreover, the dynamics governed by Schr\"odinger's equation actually conserves the overlap between two initial states.
Thus, a crucial step towards understanding how and in what sense closed quantum systems thermalize is to extend the notion of chaos and ergodicity to the quantum domain in a consistent way \cite{Haake2010, Rigol2016}. For systems which feature a well-defined classical limit it has been shown that the statistics of the energy levels of the quantum Hamiltonian are a faithful indicator of chaos in the corresponding classical model. Wigner and Dyson famously noticed that for a generic quantum many-body system the distribution of energy levels in a small energy window essentially looks like that of a random Hamiltonian matrix \cite{Wigner1993, Dyson1962}. This insight led to the conjecture that the level statistics of quantum systems that have a classically chaotic counterpart can be described by random matrix theory \cite{BGS1984} which has been confirmed numerically \cite{Brody1981,rudnick2008,Friedrich1987,Rigol2010prA,Santos2012}.

Despite these successes, the statistics of the eigenstates of a quantum Hamiltonian only provides a static picture, which limits its usefulness for answering questions about relaxation \emph{dynamics} in quantum systems. Instead, the spreading of correlations and entanglement \cite{Calabrese2009, Calabrese2005} has shifted into the focus as a tool for understanding relaxation in closed quantum systems. 
Recently, a specific type of correlation functions, so-called out-of-time-order correlators (OTOCs), have been proposed as suitable indicators of chaos in quantum systems \cite{Cotler2017,Richter2018,Larkin1969,Schuckert2019}. In particular the squared commutator $C(t)=\langle [\hat W(t),\hat V(0)]^\dagger [\hat W(t),\hat V(0)]\rangle$ \footnote{Note that sometimes only the out-of-time order part $\hat W(t)\hat V(0)\hat W(t)\hat V(0)$ is defined as OTOC, however, we will use the term OTOC also when referring to the squared commutator.} is of interest as its growth can be related to the Lyapunov exponent by heuristic semi-classical arguments. Here, $\hat V(0)$ and $\hat W(t) = \hat U(t)^\dagger \hat W \hat U(t)$ are Heisenberg operators \footnote{The name OTOC comes from the fact that upon expanding the squared commutator one obtains correlation functions in which the Heisenberg operators do not act in normal time order, i.e. cannot be put on a single Keldysh contour.}. For example, if $\hat V$ and $\hat W$ correspond to position and momentum operators $\hat x$ and $\hat p$, in the limit of small $\hbar$ the short-time behavior of $C(t)$ is obtained by replacing operators by classical fields and the commutator by the Poisson bracket (see Refs.~\cite{Cotler2017} for a more rigorous argument). This results in the quantum to classical correspondence \cite{Cotler2017, Richter2018, Rozenbaum2017, ChavezCarlos2019}
\begin{equation}
\label{eq:OTOC1}
	C(t) \rightarrow \left| \left\{x(t), p(0) \right\} \right|^2 = \left| \frac{\partial x(t)}{\partial x(0)} \right|^2 \sim  e^{2 \lambda t} \,.
\end{equation}
where $\lambda$ is the classical Lyapunov exponent. This simple argument suggests that the squared commutator grows exponentially at short times up to the Ehrenfest time \cite{Richter2018} if the corresponding classical model is chaotic.
For this reason the growth rate of the squared commutator has been termed the quantum Lyapunov exponent.
We emphasize that the mentioned semi-classical arguments only hold in systems with a well-defined classical limit of high occupation numbers. Also, as we discuss in more detail below, in higher dimensional systems the derivative appearing in Eq.~\eqref{eq:OTOC1} will, for a given OTOC, involve a specific pair of phase space coordinates [$\partial x_i(t)/\partial x_j(0)$] and thus will not necessarily grow exponentially with the largest Lyapunov exponent of the classical dynamics, which characterizes of the phase space direction of fastest growth. OTOCs have also been of interest in the context of spin systems with local interactions and random unitary circuits where they quantify operator spreading, or scrambling of quantum information. The present study focuses on the case of a bosonic few mode system where a classical mean-field description becomes exact is the limit of large particle numbers $N$.

Intense theoretical efforts have recently been undertaken to reach a more thorough understanding of the squared commutator as an indicator for quantum chaos (see \cite{Swingle2018} for a recent review). 
Measuring OTOCs experimentally has been challenging due to the requirement of implementing many-body echo protocols \cite{Gaerttner2017,Li2017,Wei2018,Niknam2018}. In particular, the predicted exponential growth of OTOCs in systems showing classical chaos in their mean-field dynamics has not been observed in experiments. This motivates us to study OTOCs in a model of three bosonic modes which can be realized experimentally by a Bose-Einstein condensate (BEC) of rubidium atoms. We demonstrate that this model shows rich dynamics being tunable between regimes of regular and chaotic dynamics and offers a way to implement the many-body echo experimentally, thus providing an ideal platform for studying OTOCs and quantum chaos.

Specifically, we consider a BEC in a tightly confining trapping potential which constrains the dynamics to the lowest trap state. The relevant degrees of freedom are the three Zeeman levels of the $F=1$ hyperfine manifold of $^{87}$Rb. The dynamics of this three-mode system is governed by mode-changing collisions between the atoms and the quadratic Zeeman shift. The classical mean field dynamics of this model is integrable and has been studied extensively including several experimental realizations \cite{Huang2015,Plimak2006,Law1998,Sengstock2005,Schmaljohann2004,Hamley2012, Klempt2018,Oberthaler2018,Linnemann2016,Gerving2012,Davis2019}.
By introducing a tunable coherent coupling term between neighboring Zeeman states \cite{Kunkel2019} integrability is broken \footnote{Recall that we refer to the integrability of classical mean field equations. Integrability is broken by a symmetry-breaking field. This is different from the often considered case of an integrable system of non-interacting particles where integrability is broken by adding interactions between the particles.}.

The objective of this work is thus to provide a detailed understanding of the classical mean field dynamics of this model and to explore signatures of classical chaos in the quantum dynamics at finite particle number $N$.
For this we first characterize the \emph{classical} dynamics by calculating Poincar\'e sections and Lyapunov exponents for varying strength of the integrability breaking term. We find that the classical phase space portraits and Lyapunov exponents show a transition from regular to mixed phase space and strongly chaotic dynamics. Turning to quantifiers of \emph{quantum} chaotic behavior we find that the level statistics changes from Poissonian to Wigner-Dyson-like (more precisely to a Brody parameter of $\approx 0.6$ \cite{Brody1981}). Finally, the squared commutator is found to generically show a faster growth for initial states located in classically chaotic regions than in regular ones. 
The short-time behavior before the Ehrenfest time, or scrambling time, $t_s\propto \log(N)$, is well described by semi-classical methods showing that the OTOC indeed reveals signatures of classical chaos in the quantum dynamics at finite $N$. We find that OTOCs with respect to an initial coherent state located in a classically chaotic region of phase space often show exponential growth at short times, however, the precise behavior is expected to depend strongly on the initial state and will, in general, not yield the largest Lyapunov exponent of the limiting classical dynamics.

Related work on OTOCs and chaos in few mode bosonic systems includes a study reporting on the growth of the squared commutator in the integrable limit of a model closely related to the one considered here \cite{Hummel2018}. The squared commutator is found to show exponential growth if the system is initially prepared at a dynamically unstable point of the classical phase space.
Classically chaotic dynamics has been found to emerge when extending this model to five modes \cite{Kronjaeger2008}, when considering the motional modes of a single component BEC in an anharmonic trapping potential \cite{Garcia2018}, and in a three-site Bose-Hubbard model (Bose trimer) \cite{Penna2003,Munro2000,Buonsante2009}. In the latter cases the interactions are quite different from the mode-changing collisions considered here, leading to different types of phase space structures. OTOCs have not yet been discussed in the case of few mode systems with classically chaotic dynamics.

The remainder of this work is structured as follows: After introducing the model Hamiltonian in Sec.~\ref{sec:model}, we  first analyse the classical phase space to determine the parameter regimes in which chaotic behaviour is encountered (Sec.~\ref{sec:classical}). Then, in Sec.~\ref{sec:quantum} we  conduct the corresponding analysis for the quantum case, looking for quantum signatures of classical chaos. Section~\ref{sec:exp} details the prospects of measuring OTOCs experimentally in a spinor BEC. We draw our conclusions and discuss possible future research directions in Sec.~\ref{sec:conclusion}.

\section{Model} \label{sec:model}
We consider the following three-mode Hamiltonian in second quantized form \cite{Gerving2012}
\begin{align} \label{eq:Hamiltonian}
	\begin{split}
	\hat H &= g \left\lbrace  (\hat a_0^{\dagger} \hat a_0^{\dagger} \hat a_1 \hat a_{-1} +  \hat a_1^{\dagger} \hat a_{-1}^{\dagger} \hat a_0 \hat a_0 ) + \hat N_0 (\hat N_1 +\hat N_{-1}) \right.\\
	&\qquad + \left. \frac{1}{2} (\hat N_1 - \hat N_{-1})^2 \right\rbrace  + q (\hat N_1 + \hat N_{-1}) \\
	&\qquad +  \frac{r}{\sqrt{2}} \left\lbrace (\hat a_1^{\dagger} + \hat a_{-1}^{\dagger} ) \hat a_0 + \hat a_0^{\dagger} ( \hat a_1 + \hat a_{-1}) \right\rbrace
	\end{split}
\end{align}
where $\hat a_i$ ($\hat a_i^{\dagger}$), with $i=0,\pm 1$, denote annihilation (creation) operators of three bosonic modes giving rise to the corresponding number operators $\hat N_i \equiv \hat a_i^{\dagger} \hat a_i $. 
In the experimental realization of this model in a spinor BEC \cite{StamperKurn2013,Gerving2012,Hamley2012, Klempt2018,Oberthaler2018,Linnemann2016} mentioned in Sec.~\ref{sec:intro} the parameter $g$ represents the strength of s-wave interactions between the atoms, $q$ is the quadratic Zeeman shift, and $r$ corresponds to a coherent coupling between the modes induced by an rf-modulation of the magnetic field. All these parameters are tunable. For more details on the proposed experimental realization see Sec.~\ref{sec:exp}.

\begin{figure*}
	\includegraphics{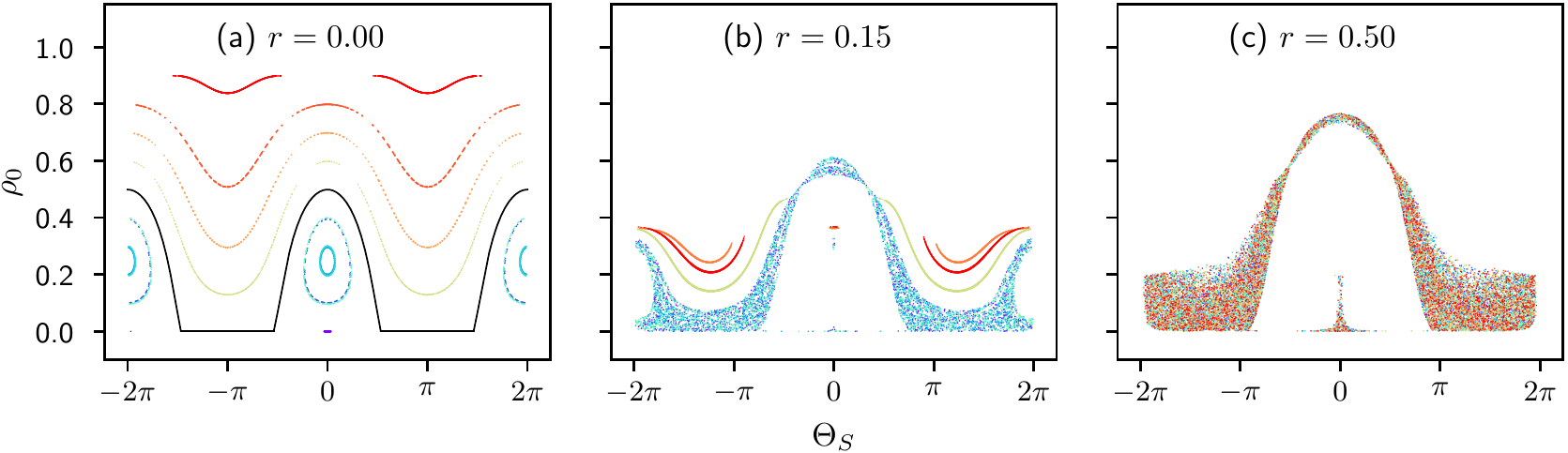} 
	\caption{Poincaré sections illustrating the phase space structure for increasing values of $r$. The plane of intersection is defined by $\varTheta_m = 0$. In (a) the points of intersection of each trajectory for a given energy are confined to a line. This changes in (b) and (c) for increasing values of $r$: In (b) we can see a mixed phase space where for some trajectories the points of intersection spread over a larger area of phase space while for others the points are still confined to lines; in (c) the phase space is strongly chaotic where all allowed regions compatible with energy and norm are reached from any initial point. 
	To be able to show Poincaré sections for different initial states in one plot, the initial states in (a) correspond to different energies. Here, the initial states are chosen by varying $\rho_0$ while keeping, $\varTheta_s = 0$, $m=0.1$ and $\varTheta_m$ fixed. The separatrix between vibrational and rotational motion is shown as a solid black line. 
	In (b) and (c) we only used initial states corresponding to an energy of $E=1.005$ chosen equal to the energy of the separatrix in the unperturbed case. To adjust the energy of the initial states for a given initial $\rho_0$ and $\varTheta_s$, we choose $m$ and $\varTheta_m$ accordingly. Different colours represent trajectories corresponding to different initial states.}
	\label{fig:poincare}
\end{figure*}

The Hamiltonian (\ref{eq:Hamiltonian}), in addition to the total energy, conserves the total number of particles $\hat N \equiv \hat N_{-1} + \hat N_0 +\hat  N_1$ and for $r=0$ it also conserves the imbalance, or magnetization, $\hat S_z \equiv \hat N_1 - \hat N_{-1}$. Thus, as we will see shortly, at $r=0$ the model is integrable in the classical mean field limit. By tuning $r \neq 0$ magnetization is no longer conserved and integrability is broken. 
In the following we fix $gN$ as the reference energy scale. Since we also set $\hbar=1$, times are given in units of $\hbar/gN$.
The observed qualitative features are independent of the exact choice of $q/gN$ as long as it is at most of order $1$. We thus fix $gN = q = 1$ in the following discussion. 
The parameter $r$ will be varied from the classically integrable case ($r=0$) to the strongly chaotic case ($r\approx 1$).

Bosonic systems with a finite number of modes can be described by classical mean-field equations in the limit of large mode occupations and at times shorter than the Ehrenfest time \cite{Sciolla2011}, which scales as $\sqrt{N}$ for integrable and $\log(N)$ for classically chaotic systems. For finite $N$ the dynamics can be described as that of classical wave packets of width $\propto 1/\sqrt{N}$ and thus the classical limit is controlled by $\hbar\ind{eff}\sim 1/N$ \cite{Pappalardi2018}.
Technically, the classical mean field Hamiltonian is obtained by replacing the creation and annihilation operators $\hat{a}_i$ and $\hat{a}^{\dagger}_i$ by $\zeta_i\sqrt{N}$ and $\zeta_i^{\star}\sqrt{N}$, respectively, where $\zeta_i$ are complex-valued, normalized classical fields. This approximation becomes exact in the large-$N$ limit where terms of order $1/N$ are negigible \cite{Gardiner2004}.
The resulting mean-field Hamiltonian reads (keeping $gN$ fixed)
\begin{align} \label{eq:Hmf_zeta}
    \begin{split}
    	H\ind{mf} &= g N \left\lbrace \zeta_0^{\ast 2 } \zeta_1 \zeta_{-1} + \zeta_1^{\ast} \zeta_{-1}^{\ast} \zeta_0^2 +  |\zeta_0|^2 \left( |\zeta_1|^2 + |\zeta_{-1}|^2 \right) \right. \\   
    	&\quad \left. +  m^2 / 2 \right\rbrace +  q \left( |\zeta_1|^2 + |\zeta_{-1}|^2 \right)	\\
    	&\quad + r / \sqrt{2} \left\lbrace (\zeta_1^{\ast} + \zeta_{-1}^{\ast} ) \zeta_0 + \zeta_{0}^{\ast} (  \zeta_1 + \zeta_{-1}) \right\rbrace \ .
    \end{split}
\end{align}
From Eq.~\eqref{eq:Hmf_zeta} one straightforwardly derives the equations of motion (setting $\hbar = 1$)
    \begin{align} \label{eq:MFeom}
	\begin{split}
		i\! \ \dot{\zeta}_1 &= q \ \zeta_{1} + g N \left\lbrace ( \rho_1 + \rho_0 - \rho_{-1} ) \zeta_{1} +  \zeta_{-1}^{\star} \zeta_0^2   \right\rbrace 
		+ \frac{ r }{\sqrt{2}} \ \zeta_0 \\
		i\! \ \dot{\zeta}_0 &=  g N \left\lbrace ( \rho_1 + \rho_{-1} ) \zeta_{0} +  2 \zeta_{0}^{\star} \zeta_1 \zeta_{-1}  \right\rbrace 
		+ \frac{ r }{\sqrt{2}} \ (\zeta_1 + \zeta_{-1} ) \\
		i\! \ \dot{\zeta}_{-1} &= q \ \zeta_{-1} + g N \left\lbrace ( - \rho_1 + \rho_0 + \rho_{-1} ) \zeta_{-1} +  \zeta_{1}^{\star} \zeta_0^2   \right\rbrace \\
		& \hspace{6.4cm} + \frac{ r }{\sqrt{2}} \ \zeta_0 
	\end{split}
\end{align}
where $\rho_i=|\zeta_i|^2$. We integrate these equations numerically in order to find the classical time evolution.

It is convenient to reparametrize the fields $\zeta_i$ by explicitly exploiting conservation laws \cite{Gerving2012}: We express the fields by real amplitudes and phases $\zeta_j = \sqrt{\rho_j} \ e^{i \varTheta_j}$.
As mentioned above, the Hamiltonian conserves the total particle number $N$ which implies that the norm of $\sum_i \rho_i$ is constant ($=1$). Fixing also the global phase 
reduces the number of relevant classical variables (phase space coordinates) from six to four \cite{Hamley2012}
\begin{align} \label{eq:paramZeta}
	\begin{split}
		\zeta_1 &= \sqrt{\frac{1-\rho_0 + m}{2}} \ e^{i \frac{\varTheta_s + \varTheta_m}{2}} \\
		\zeta_{0} &= \sqrt{\rho_0} \\
		\zeta_{-1} &= \sqrt{\frac{1-\rho_0 - m}{2}} \ e^{i \frac{\varTheta_s - \varTheta_m}{2}}, 
	\end{split}
\end{align} 
with the normalized magnetization $m = \rho_1 - \rho_{-1} = (N_1 - N_{-1})/N$, the Larmor precession phase $\varTheta_m = \varTheta_1 - \varTheta_{-1}$ and the spinor phase $\varTheta_s = \varTheta_1+ \varTheta_{-1} - 2 \varTheta_0$. 
As we will use this parametrization in the following description of the classical phase space, it is useful to rewrite the mean field Hamiltonian as

\begin{align} \label{eq:Hmf}
	\begin{split}
		H\ind{mf} &= g N \rho_0 \left\lbrace \left( 1 - \rho_0 \right) + \sqrt{\left( 1 - \rho_0 \right)^2 - m^2 } \ \cos \varTheta_s \right\rbrace \\
		& \quad + \frac{g N}{2} m^2 + q \left( 1 - \rho_0 \right) \\
		& \quad + r \sqrt{\rho_0} \left\lbrace \sqrt{1-\rho_0 + m} \ \cos \left(  \frac{\varTheta_s + \varTheta_m}{2} \right) \right.\\
		&\qquad \left. + \sqrt{1-\rho_0 - m} \ \cos \left(  \frac{\varTheta_s - \varTheta_m}{2} \right)  \right\rbrace \,.
	\end{split}
\end{align}
As the canonical transformation (\ref{eq:paramZeta}) leaves us with two pairs of canonical coordinates $\left\lbrace \rho_0, \varTheta_S \right\rbrace$ and $\left\lbrace m, \varTheta_m \right\rbrace$, we immediately see from Eq.~(\ref{eq:Hmf}) that the magnetization $m$ is conserved for $r=0$. 

\section{Classical phase space} \label{sec:classical}
In this section we characterize the classical dynamics and study its dependence on the strength of the integrability breaking term. We identify regimes of classically chaotic dynamics adopting the following notion of classical chaos. For a system to be chaotic we require that it is both sensitive to initial conditions, meaning that two initially close-by trajectories diverge exponentially, and also that any two trajectories with equal energy and possibly other conserved quantities come arbitrarily close to each other at later times. While the first requirement can be checked numerically by calculating Lyapunov exponents, the second is implied by ergodicity (almost every trajectory passes arbitrarily close to almost every point in phase space) \cite{Zirnbauer1988} which can be numerically verified by studying Poincaré sections.

\subsection{Poincaré sections}
Instead of looking at the full four dimensional trajectories in classical phase space, we only record their crossings with a certain (three dimensional) surface of section, here defined by $\varTheta_m = 0$. For visualization, we further use that in a time-independent classical Hamiltonian system energy is conserved, such that we can project our Poincaré section onto a two dimensional plane (in this case the $\left\lbrace \rho_0, \varTheta_s\right\rbrace $-plane, see Fig.~\ref{fig:poincare}). 

The Poincaré sections contain all relevant information to determine the range of parameters and initial states in which the system exhibits ergodic behaviour \cite{berry1978}.
When the intersection points are strictly confined to a one-dimensional subspace of the section, dynamics is regular. When the points of intersection are instead spread over a larger area, only limited by conservation of energy and norm, the system is likely to be ergodic in the sense described above.
In a chaotic system the described phase space properties can be found for arbitrary surfaces of section but it turns out that the choice of the hyperplane influences how well the transition from regular to chaotic behaviour can be visualized. We found the surface defined by $\varTheta_m = 0$ to be a suitable choice in the case at hand.

Figure~\ref{fig:poincare}(a) shows a sample of trajectories for the regular case $r=0$. In this case $m$ is conserved and thus $\theta_m$ is cyclic. To be able to show multiple Poincaré sections in one plot, each initial state (colour encoded) corresponds to a different energy. The  Poincaré sections in this regular case are confined to one dimension (lines). 
Figures~\ref{fig:poincare}(b) and (c) show the Poincaré sections for non-integrable cases. 
We used a raster of initial states in the $\left\lbrace \rho_0, \varTheta_s\right\rbrace $-plane with $m$ and $\varTheta_m$ chosen such that all trajectories have the same fixed energy of $E = 1.005$. This energy shell was chosen in order to ensure that the energetically allowed region in the $\left\lbrace \rho_0, \varTheta_s\right\rbrace $-plane stays sufficiently large for all considered values of $r$. It matches the energy of the separatrix between vibrational and rotational motion [solid black line in Fig.~\ref{fig:poincare}(a)] in the integrable case ($r=0$).

We observe the transition from a regular [Fig.~\ref{fig:poincare}(a)] to a strongly chaotic phase space [Fig.~\ref{fig:poincare}(c)], transitioning a mixed phase space [Fig.~\ref{fig:poincare}(b)] where depending on the initial state, a trajectory is confined to a one dimensional subspace or samples a larger part of the available phase space. The regular structure first starts to dissolve around the separatrix. The regions around the stable fixed points of the integrable dynamics retain closed orbits up to relatively large values of $r$. In the case of $r=0.5$, where we do not observe any regular islands any more in the shown energy shell, we still found mixed phase space structure at other energies, which is why we do not call this case fully chaotic but rather strongly chaotic.

We note here that we computed Poincaré sections for a large set of values of $q$ and $r$ (and accordingly adjusted energy $E$) in addition to the ones shown in Fig.~\ref{fig:poincare} and found that the qualitative features are the same in all cases: If one of the parameters dominates (e.g. $q\gg 1$ or $r\gg 1$), or if $r\approx 0$, the system looks regular. If all parameters are of similar magnitude, ergodic parts of the phase space emerge and eventually dominate. This transition is not as well visible for all choices of parameters and energy since regions where regular and chaotic patches coexist may only be found in certain parts of the phase space.

\begin{figure*}
	\includegraphics{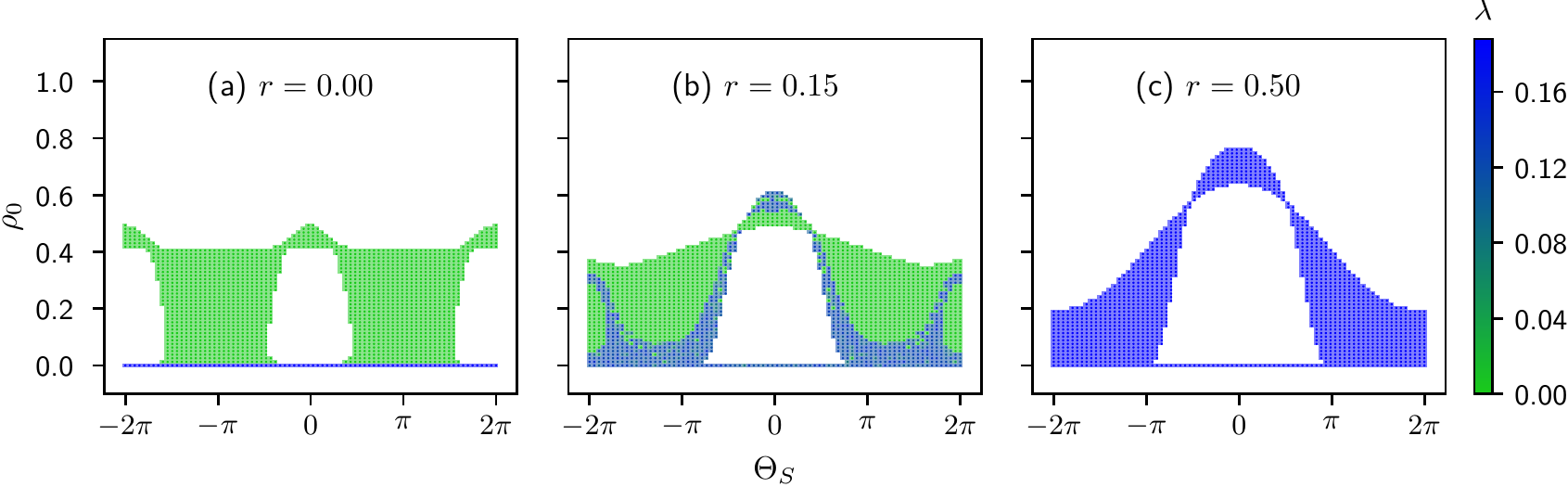}
	\caption{Lyapunov exponents $\lambda$ (colour encoded) for a raster of $80 \times 80$ values of $\rho_0$ and $\varTheta_s$, $\varTheta_m = 0$. At each point the value of $m$ has been adjusted in order to match a given energy, here $E=1.005$. The choice of energy mainly influences the allowed compatible regions in phase space but not the qualitative features of the phase space portrait. The plots are shown for increasing values of $r$. As in Fig.~\ref{fig:poincare}, we see the transition from a regular (a) through a mixed (b) to a strongly chaotic (c) phase space. }
	\label{fig:lyapunov}
\end{figure*}

\subsection{Lyapunov exponents} \label{subs:LyapunovExp}
As an ergodic phase space is only a necessary condition for classical chaos, we also need to check for the linear instability of the trajectories, measured by the Lyapunov exponent.
The sensitivity of a dynamical system to initial conditions can be quantified by the (largest) Lyapunov exponent $\lambda$ \cite{Cotler2017}. Given a classical trajectory in phase space $\vec x(t)$ and some initial conditions $\vec x_0$, the sensitivity to initial conditions can be expressed as the exponential divergence of initially infinitesimally separated trajectories:
\begin{equation}
	\frac{\left|\Delta \vec x(t)\right|}{\left|\Delta \vec x_0\right|}  \sim e^{\lambda t} \,.
\end{equation}
Thus, the largest Lyapunov exponent $\lambda$ is defined as
\begin{equation} \label{eq:LE}
    \lambda = \lim_{t \to \infty}\lim_{\left|\Delta \vec x_0\right| \to 0} \frac{1}{t} \log \frac{\left|\Delta \vec x(t)\right|}{\left|\Delta \vec x_0\right|} \,.
\end{equation}
For chaotic systems the largest Lyapunov exponent $\lambda$ is greater than zero, i.e. we are interested in the long-time limit as this will converge to the largest Lyapunov exponent for almost all orientations of the initial separation \cite{Eckmann2004}.

For the numerical calculations we use an equivalent definition of the Lyapunov exponent explicitly in terms of two (infinitesimally) close trajectories $\vec{z}_1(t)$ and $\vec{z}_2(t) \equiv \vec{z}_1(t) + \vec{\xi}(t)$. Then the largest Lyapunov exponent $\lambda$ can be obtained as (see e.g. \cite{Wolf1985})
\begin{equation} \label{eq:LE_xi}
	\lambda = \lim_{t \to \infty} \frac{1}{t} \log \frac{\left| \vec{\xi}(t)  \right|}{\left| \vec{\xi}(0) \right|} \ .
\end{equation}
For evaluating this quantity we employed the following numerical procedure \cite{Tarkhov2017}: Given a starting point in phase space $\vec{z}_1(t_0)$ we choose a random point $\vec{z}_2(t_0)$ with initial distance $\xi_0=\left| \vec{\xi}(0) \right|$ from $\vec{z}_1(t_0)$. We then calculate the time evolution for both trajectories simultaneously by numerical integration. After each time step $l(t) = \log(\xi(t)/\xi_0)$ is evaluated and recorded, where $\xi(t) = \left| \vec{z}_1(t) - \vec{z}_2(t) \right|$. When reaching the reset time $T_r$ the current Lyapunov exponent is calculated as $\lambda_t = l(T_r)/ T_r$ and the distance between the trajectories is reset to $\xi_0$ by shifting $\vec{z}_2(t)$ to $\vec{z}_2^{\prime}$ along the vector $\vec{z}_1(t)-\vec{z}_2(t)$. This procedure is repeated up to a certain time and the largest Lyapunov exponent is calculated as the average over all $\lambda_t$ omitting the values up to a time $t_{\mathrm{min}}$. We omit these first values to ensure that we have reached the basin of attraction which is important to get a sensible estimate of the Lyapunov exponent. An alternative method which is inherently linear in the initial separation is evolving the fundamental matrix $\Phi_t(\vec{\zeta})$ together with the equations of motion $\dot{\vec{\zeta}} = \vec{F}(\vec{\zeta})$ \cite{Skokos2010}:
\begin{equation}
    \left(\begin{matrix} \dot{\vec{\zeta}} \\
    \dot{\Phi} \end{matrix} \right) = 
    \left(\begin{matrix} \vec{F}(\vec{\zeta}) \\
    D_{\vec{\zeta}} \vec{F}(\vec{\zeta}) \ \Phi(\vec{\zeta}) \end{matrix} \right) \quad .
\end{equation}
Here, $D_{\vec{\zeta}} \vec{F}(\vec{\zeta})$ denotes the Jacobi matrix of the equations of motion. To ensure differentiability, we work in a (real) six dimensional phase space consisting of the real and imaginary parts $\zeta_i^{R,I}$, $i \in \lbrace-1,0,1 \rbrace$ of the components of $\vec{\zeta}$. At $t=0$ we set $\vec{\zeta}$ according to the desired initial state and $\Phi = \mathbb{1}_{6x6}$. Having calculated the time evolution of the fundamental matrix $\Phi_t$, which contains derivatives of the form $\partial \zeta_i^{R/I}(t) / \partial \zeta_j^{R/I}$ evaluated along the trajectory $\vec{\zeta}(t)$, we obtain the deviation vector at time $t$ as $\vec{\xi}(t) = \Phi_t \vec{\xi}(0)$.
Choosing again a random initial deviation vector $\vec{\xi}(0)$, this provides an alternative way to calculate the largest Lyapunov exponent via eq. \eqref{eq:LE_xi}. We checked that both methods yield consistent results within numerical errors. 

In Fig.~\ref{fig:lyapunov} we show the Lyapunov exponents calculated as described above for a raster of initial points in the $\left\lbrace\rho_0, \varTheta_s\right\rbrace$-plane corresponding to the same energy used to generate the Poincaré sections in Fig.~\ref{fig:poincare}. 
We find the same qualitative signatures of classical chaos as for the Poincaré sections in Fig.~\ref{fig:poincare}: The integrable system ($r=0$) shows no signs of classical chaos. Increasing $r$, we first observe a mixed phase space where it depends on the initial state whether a trajectory has zero or positive Lyapnuov exponent and thus whether it exhibits chaotic behaviour. Finally, for a perturbation $r$ of the same order as the other contributions to the Hamiltonian (\ref{eq:Hamiltonian}), we see that chaotic regions clearly dominate the phase space. As for the Poincaré sections, the same qualitative features are found for other combinations of the parameters $q$ and $r$. 

\section{Quantum signatures of chaos} \label{sec:quantum}
Our aim in this section is to reveal features of the quantum system that can be traced back to chaos in the corresponding classical system and thus may be useful indicators of ergodicity in the quantum system.
For this we solve the model \eqref{eq:Hamiltonian} by exact diagonalization in Fock space for $N\lesssim 100$, which is feasible as the Hilbert space dimension scales as $N^2$.

\subsection{Level statistics}
A well-established indicator for chaos -- or, more precisely, for randomness -- in quantum mechanical systems is the distribution of the energy eigenvalue spacings of the Hamiltonian \cite{mehta2004}. 

If a regular system is perturbed, formerly allowed level crossings (e.g.\ due to symmetries in the unperturbed system) will be avoided. This level repulsion changes the distribution of energy level spacings: In the unperturbed case, as level crossings are allowed, the distribution is peaked at zero and follows a Poisson distribution $P(s) = e^{-s}$. Now with level repulsion, small spacings are suppressed and the distribution of energy level spacings is best approximated by a Wigner distribution $P(s) = \pi/2 \ s \ e^{-\pi s^2 /4}$, which is what is also obtained for random matrices \cite{Brody1981, Grass2013}.

\begin{figure}
	\includegraphics{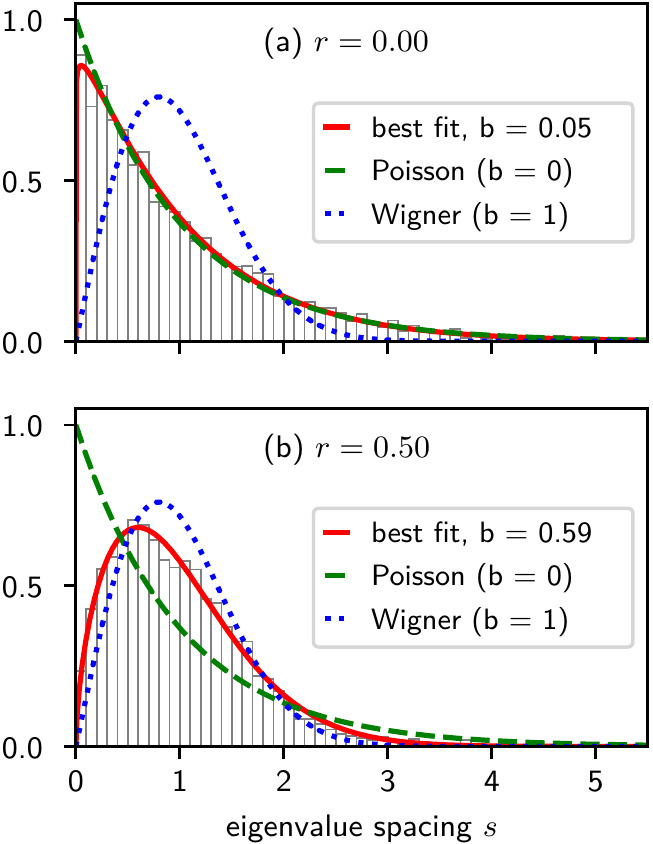} 
	\caption{Histograms of the level spacing distributions for different values of $r$. We show the combined distributions of the subspaces defined by the spin flip symmetry, see text. $N=100$ atoms, leading to 5091 energy eigenvalues, have been used. The solid red lines are fits of a Brody distribution to the data.}
	\label{fig:levelstat}
\end{figure}

\begin{figure*} 
	\hspace{-0.75cm}\includegraphics{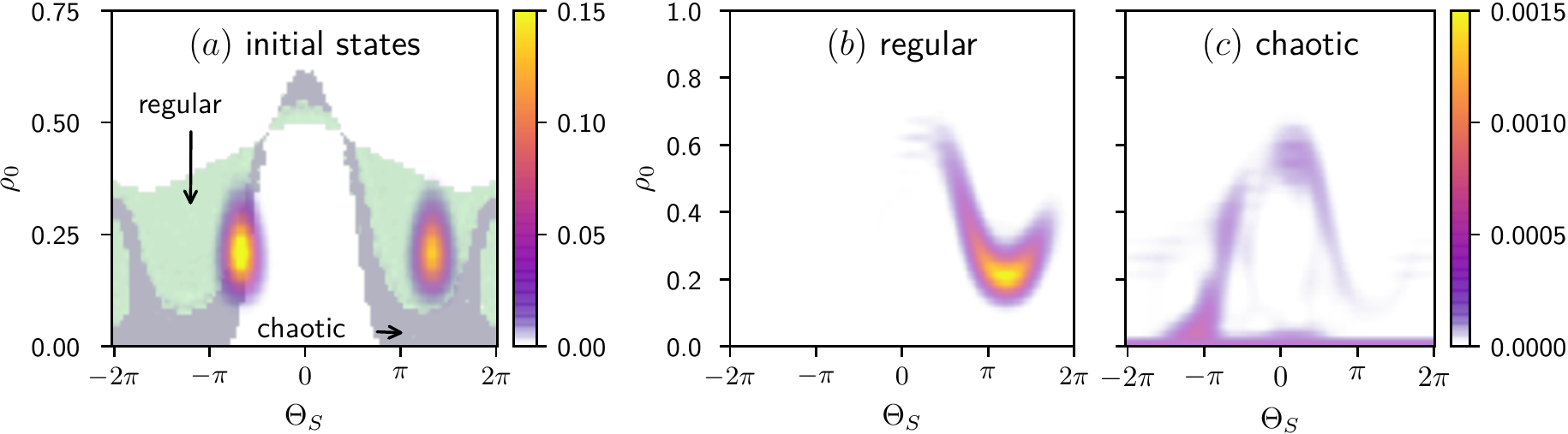}
	\caption{(a) Husimi distributions of the two initial coherent states (colour encoded) plotted on top of the outline of the classical Lyapunov exponent for $r = 0.15$ (cf.\ Fig.~\ref{fig:lyapunov}). The state on the right hand side corresponds to the coherent state centered at $\rho_0 = 0.2$, $\varTheta_s = 4\pi/3$, $\varTheta_m = 0$ and $m$ chosen such that its energy corresponds to $E=\langle \hat{H} \rangle=1.005$ which lies in the classically regular region of the phase space for the given parameters. The left one is centered in the classically chaotic region at $\rho_0 = 0.2$, $\varTheta_s = - 2\pi/3$, $\varTheta_m = 0$ and $m$ chosen again such that its energy corresponds to $E=1.005$.  Figures (b) and (c) show the Husimi distributions of the time-evolved states at $t = 10$ corresponding to both initial states in (a).}
	\label{fig:CoherentStates}
\end{figure*}

As the Hamiltonian (\ref{eq:Hamiltonian}) possesses a $\mathbb{Z}_2$ symmetry consisting in the invariance under exchange of the state labels $1$ and $-1$, i.e. under $m \mapsto -m$, we first transform to an eigenbasis of this symmetry such that the Hamiltonian matrix becomes block diagonal and calculate the spectrum for each block separately. 
In addition, we unfold the spectrum in each block to remove the influence of the level density (see e.g. \cite{Gubin2012}) before finally recombining the statistics from the different blocks. The resulting distribution of level spacings $s$ is visualised in Fig.~\ref{fig:levelstat} for the values $r=0$ and $r=0.5$ where also the Poisson (dashed green) and Wigner (dotted blue) distributions are shown. To get a quantitative estimate of how strongly chaotic the Hamiltonian is we fit the so-called Brody distribution $P_b(s)$ to the distributions (solid red). This function interpolates between the Poisson ($b=0$) and the Wigner distribution ($b=1$) and is given by \cite{Brody1981}
\begin{equation}
P_b(s) =  \alpha (b+1) s^b \exp \left[ -\alpha s^{b+1} \right]  \ , \
\alpha = \left[ \Gamma \left( \frac{b+2}{b+1} \right) \right]^{b+1} \,.
\end{equation}

In Fig.~\ref{fig:levelstat} we see that level statistics nicely reflect the classical analysis: For $r=0$ we essentially get a Poisson distribution and for $r = 0.5$ the Brody parameter $b\approx 0.6$ suggests that the system is strongly chaotic as expected from the analysis of the classical phase space. The observed $b<1$ is most likely due to the presence of islands of stability at other energies than the one shown in Fig.~\ref{fig:poincare}(c). The level spacing distributions shown in Fig.~\ref{fig:levelstat} are averaged over the full spectrum. A more thorough analysis would require an energy resolved analysis of the spectral statistics, for which larger Hilbert space dimension would be necessary in order to keep statistical fluctuations small.

A drawback of level statistics as an indicator of quantum chaos is that they only reflect global properties (independent of initial states) and thus do not allow to distinguish the relaxation dynamics of initial states corresponding to different regions of the classical phase space. Also, they are hardly accessible experimentally, which motivates us to turn to experimentally measurable quantities that allow us to resolve the phase space structure, namely Husimi distributions and OTOCs, in the following.

\subsection{Dynamics of Husimi distributions}

Husimi distributions allow the visualization of quantum states in a quantum phase space, which is convenient for drawing analogies to the classical picture. In the following, we focus on the case of a classically mixed phase space ($r=0.15$) and consider as initial states coherent states located in the classically chaotic and regular region, respectively, see Fig.~\ref{fig:CoherentStates}(a).
In order to get an intuition for the differences between these two cases, we study the evolution of the Husimi distribution of each state. The Husimi distribution of a state $\ket{\psi}$ is a phase space representation and is defined as 
$$ 
Q(\alpha) = \left| \langle {\alpha} \ket{\psi} \right|^2 
$$ 
for coherent states $\lbrace \ket{\alpha}\rbrace$.
To obtain the Husimi distribution of a state $\ket{\psi}$ numerically we would need to calculate the overlap of $\ket{\psi}$ with each element of a four dimensional grid of coherent states $\lbrace \ket{\rho_0, \varTheta_s, m, \varTheta_m} \rbrace$.
But as we ultimately want to obtain a two-dimensional phase space representation of $\ket{\psi}$ that we can compare to the Poincaré sections (see Sec. \ref{sec:classical}), we fix $\varTheta_m = 0$ and, for each value of $\rho_0$ and $\varTheta_s$, sum the projections over a grid of $m$-values. The Husimi distributions of two initial coherent states and their time evolution are shown in Fig.~\ref{fig:CoherentStates}.

In the limit of large $N$ the size of a coherent state represented in this way would shrink to zero as $1/\sqrt{N}$. Here we use a moderate value of $N=100$ where the extension of the distribution in phase space is still large compared to the classical phase space features we wish to resolve. In fact, the width, i.e. the quantum fluctuations, are so large that both initial states overlap with classically regular as well as chaotic regions of phase space. For the following discussion we should keep in mind that this quantum mechanical smearing over the classical phase space is what fundamentally limits the applicability of  semi-classical methods and leads to their breakdown at long times. This point is further discussed in Sec.~\ref{sec:OTOCs} and in the appendix.

Figures~\ref{fig:CoherentStates}(b) and (c) show the Husimi distributions of the time-evolved state $\ket{\psi(t)}$ at time $t=10$. We find that despite the relatively large initial fluctuations qualitative differences between the two cases are clearly visible. The 'regular' state remains compact in phase space and still resembles a well-localized wave packet while the 'chaotic' state is distributed across almost the whole classically allowed phase space. This confirms the expectation that the Ehrenfest time, i.e. the time at which quantum interference effects become relevant and the semi-classical wave-packet description breaks down \cite{Richter2018,  Hummel2018, ChavezCarlos2019, Schubert2012, Tomsovic2018, Kurchan2018, Scaffidi2017}, is much shorter for chaotic ($\sim \log(N)$) than for regular dynamics ($\sqrt{N}$) \cite{Pappalardi2018}.

\subsection{OTOCs}
\label{sec:OTOCs}

\begin{figure*} 
	\includegraphics[width=\textwidth]{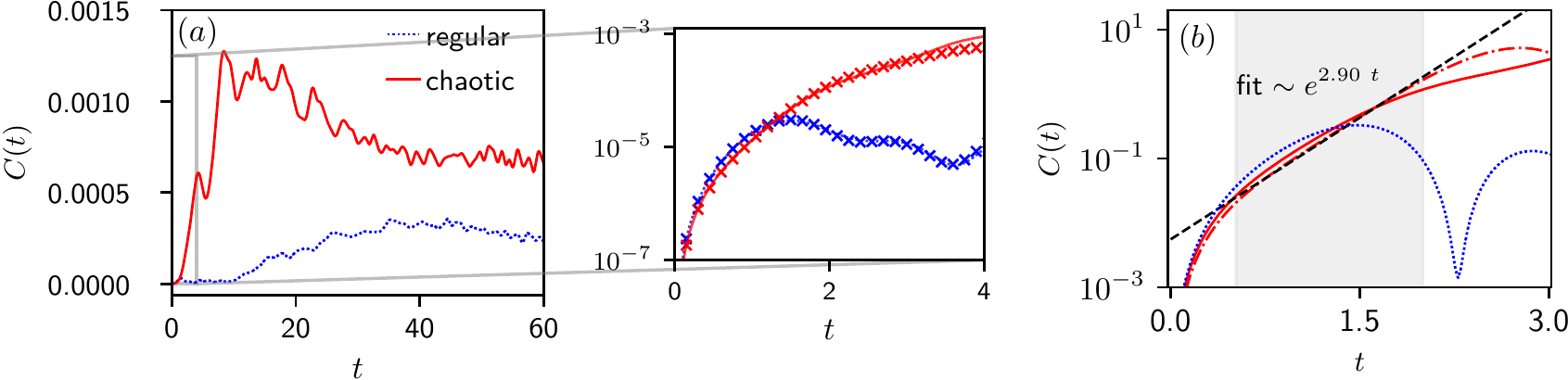}
	\caption{(a) Time evolution of the OTOC $C(t) = \left\langle \psi_0 \left| \left| \left[ \hat{\rho}_0(t),\hat{\rho}_0 \right] \right|^2 \right| \psi_0 \right\rangle$ for the two initial coherent states $\ket{\psi_0}$ shown in Fig.~\ref{fig:CoherentStates}(a) for $N=100$ particles (dotted blue line for classically regular state, solid red line for classically chaotic state) from the exact quantum mechanical calculations. The inset shows the short-time behaviour (quantum calculation marked as crosses in the respective colour) where the appropriately scaled OTOCs calculated using TWA are shown as dotted blue and solid red lines, respectively. Note, that around $t \gtrsim 3.5$ the TWA starts to deviate from the exact quantum mechanical solution for the classically chaotic initial state. (b) OTOCs $C(t)$ calculated using TWA for the same parameters as in (a) and Fig.~\ref{fig:CoherentStates}. Two OTOCs are shown in red for the classically chaotic initial state; for $N = 100$ particles (solid) and $N = 10^4$ particles (dash-dotted). We also show an exponential fit for the OTOC with $N=10^4$ as a black dashed line using the data in the shaded area. The OTOC for the classically regular initial state also for $N=10^4$ is shown as a dotted blue line to illustrate that at sufficiently high particle numbers the OTOC can indeed be used for a proxy of chaotic behaviour in the limiting classical dynamics.}
	\label{fig:OTOC}
\end{figure*}

As discussed in Sec.~\ref{sec:intro}, semi-classical arguments suggest that OTOCs of the form $C(t)=\langle [\hat W(t),\hat V(0)]^\dagger [\hat W(t),\hat V(0)]\rangle$ should grow exponentially at short times for systems that exhibit chaotic dynamics in the classical limit. This exponential regime is preceded by an initial powerlaw growth up to the so-called dephasing time and extends up to the Ehrenfest time, or scrambling time. The extension of the window of exponential growth in general depends on the chosen observables and initial states \cite{Pappalardi2019}. In the following we discuss to what extent this picture applies to OTOCs with respect to initial coherent states in the three-mode model under study. For models with collective interactions \cite{Rozenbaum2017, Schmitt2019, ChavezCarlos2019, LewisSwan2019, Marino2019, Sieberer2018} the classical limit is well-defined and corresponds to the limit of large particle number. We note that OTOCs have also been studied for systems with local interactions and low-dimensional local degrees of freedom, like spin systems. There, no meaningful classical limit exists and exponential growth is generally not expected \cite{Nahum2018, Keyserlingk2018, Xu2018a, Khemani2018}. Nevertheless, in this context OTOCs have proven to be useful measures for the spatial spreading of correlations \cite{Hosur2016, Bohrdt2017, Roberts2016} or diagnosing many-body localization \cite{Huang2017, Swingle2017}, to give some examples. Here, instead, we focus on a system of \emph{collectively} interacting bosons and ask whether signatures of the limiting classical dynamics can be observed at finite $N$.

We discuss the time evolution of the squared commutator specifically focusing on the case of $\hat V = \hat W =\hat \rho_0 \equiv \hat N_0/N$, and keeping the same parameters and initial coherent states as in the previous section.
The evolution of $C(t)$ is shown in Fig.~\ref{fig:OTOC}(a).
The OTOC of the initial state localized in a regular region remains small initially, which is consistent with the regular state evolving quasi-periodically for a long time as observed for the evolution of the Husimi distribution. The OTOC of the chaotic initial state rises extremely fast in comparison.
At short times, the onset of the expected exponential behavior is observed in the chaotic case while it is absent in the regular case (inset). At late times, corresponding to the Heisenberg or recurrence time given by the typical inverse level spacing \cite{Richter2018, Kurchan2018}, the OTOCs reach a saturation value in both cases.

Next, we study the short-time dynamics (before the Ehrenfest time) in more detail. In this regime we can employ the semi-classical truncated Wigner approximation (TWA), allowing us to simulate arbitrarily large $N$. The squared commutator can be evaluated in TWA through a phase space average over the Poisson bracket \cite{ChavezCarlos2019,LewisSwan2019, Schuckert2019, Pappalardi2019, Schmitt2019, Rozenbaum2017, Cotler2017} (see the appendix \ref{app_sec:TWA} for details)
\begin{equation}
\label{eq:OTOC_TWA}
    \left\langle\left| \left[ \hat{\rho}_0(t), \hat{\rho}_0(0) \right] \right|^2\right\rangle
    = \hbar\ind{eff}^2 \left\langle \left| 2 \zeta_0^R(0) \frac{\partial \rho_0(t)}{\partial \zeta_0^I} \right|^2 \right\rangle_W
\end{equation}
where $\langle \dots \rangle_W$ denotes the average over the Wigner function of the initial coherent state. The above expression is based on an expansion in orders of the size of phase space cells $\hbar\ind{eff}\sim 1/N$ to order $\hbar\ind{eff}^2$. 
We first validate this approach by comparing to exact diagonalization results for $N=100$ in the inset of Fig.~\ref{fig:OTOC}(a). This shows that while in the regular case, the OTOC is reproduced perfectly up to rather long times, it starts to deviate after the short-time growth regime in the chaotic case. The short-time behavior is always captured well. We note that we had to rescale the TWA results by a factor of order $1$, which depends on the initial state and Hamiltonian parameters but not on the atom number $N$, in order to match the exact diagonalization result. Further benchmark simulations including single-time observables are provided in appendix \ref{app_sec:TWA} showing agreement of TWA with exact diagonalization to much longer times than observed for the OTOCs. We can thus confidently use TWA to explore the short-time dynamics in the large $N$ regime.

As can be seen from Fig.~\ref{fig:OTOC}(b), particle numbers of $N \gtrsim 10^4$ show a more pronounced exponential growth regime at short times. However, this regime does not extend to longer times if we increase the particle number further.
Given the observation of exponential growth in the dynamics of the OTOC one may now want to compare the growth rate to the classical Lyapunov exponents shown in Fig.~\ref{fig:lyapunov}. Extracting the growth rate by an exponential fit in Fig.~\ref{fig:OTOC}(b) we found large deviations from the classical Lyapunov exponent.
We emphasize that an agreement between the two is not expected in general. As follows from Eq.~\eqref{eq:OTOC_TWA}, the specific OTOC we consider probes the stability in specific phase space directions ($\zeta_0^R$ and $\zeta_0^I$) with respect to initial changes in another direction ($\zeta_0^I$), see Appendix~\ref{app_sec:TWA} for details. This corresponds to two specific elements of the fundamental matrix $\Phi$, while the largest Lyapunov exponent is related to the largest eigenvalue of this matrix. As described in Ref.~\cite{Tarkhov2017}, a derivative such as $\partial \zeta_0^R(t)/\partial \zeta_0^I(0)$ evaluated with respect to a single initial condition will in general show erratic behavior including oscillations and at most its envelope is expected to grow exponentially. Exponential growth with the largest Lyapunov exponent $\lambda$ can only be expected after averaging over all initial states of an energy shell. This averaging is what is accomplished by the re-set procedure we use in Sec.~\ref{sec:classical} for calculating the Lyapunov exponents. 
In conclusion, for systems with higher dimensional phase spaces ($d>2$), OTOCs of specific observables and initial coherent states may show clear differences between chaotic and regular initial states but in general do not reveal the largest Lyapunov exponent of the limiting classical dynamics.

We also note that OTOCs of initially commuting operators always show powerlaw growth at short times \cite{Pappalardi2019}. The discussions of quantum vs.\ classical Lyapunov exponents one finds in the literature typically consider cases where the squared commutator is finite at $t=0$ allowing exponential growth to start immediately \cite{Prakash2019, Khemani2018, Rozenbaum2017, Schmitt2019, Stanford2016}.


We calculated $C(t)$ for a large range of other initial conditions and choices of the parameters $q$ and $r$ and evaluated a variety of other OTOCs, for example using spin operators such as $\hat S_z$ and $\hat S_x = [\hat a_0^\dagger (\hat a_1+\hat a_{-1})+\rm{h.c.}]/\sqrt{2}$. While the overall features of initial growth and saturation at long times are the same in all cases, we cannot always clearly identify exponential growth at short times for classically chaotic initial states which we attribute to the fact that we probe specific elements of the stability matrix, which may show arbitrary behavior in general \cite{Tarkhov2017}, and that the time window between the initial powerlaw growth and the Ehrenfest time may be small. For small values of $r$ we consistently found that if the initial state overlaps with the position of the separatrix of the (regular) classical phase space, the OTOC $C(t)$ grows much faster than if it lies close to the stable fixed points.
We conclude that despite OTOC growth not being directly related to the largest Lyapunov exponent of the corresponding classical dynamics, 
OTOCs show clear differences between initial states localized in classically chaotic and regular regions of phase space, thus qualifying as indicators of quantum chaos.

In order to simulate dynamics past the Ehrenfest time, and in particular to study the asymptotic behavior of OTOCs in the long-time regime we need to resort to exact diagonalization. An exciting alternative is to measure OTOCs in quantum simulation experiments with spinor condensates, discussed in the following, where particle numbers up to $10^4$ are reached easily.

\section{Experimental realization}
\label{sec:exp}

In this section we discuss a protocol for measuring OTOCs experimentally based on time reversing the system dynamics and propose an implementation using a spinor BEC of rubidium atoms.

The most intuitive protocol for testing reversibility and chaos in a quantum system is the following: Prepare an initial state, evolve it to time $t$, apply a perturbation $\hat W$, evolve for another period $t$ under the sign-reversed Hamiltonian, and then measure an observable $\hat V$. This protocol results in the measurement of $\langle \hat W(t)^\dagger \hat V \hat W(t) \rangle$. If the initial state is an eigenstate of $\hat V$, then the measured quantity has the desired form of the out-of-time-order part of the squared commutator \cite{Gaerttner2017}. Note, however, that in this protocol $\hat W$ is the unitary operator that induces the applied perturbation. In order to make the connection to Lyapunov exponents we need $\hat W$ to be a Hermitian operator. For spin-$1/2$ chains one can for example choose $\hat W$ to be a spin flip on one of the spins, $\hat \sigma_i^x$, which is Hermitian (and unitary). In the case of bosonic systems, however, this simplification cannot be made.

A possible ansatz for solving this problem is to consider small perturbations and to extract the resulting signal to low order in the perturbation strength. Mathematically, we consider $\hat W(0)=\exp[-i\phi \hat A]$, with a Hermitian operator $\hat A$. To second order in $\phi$ one obtains
\begin{equation}
\label{eq:phi_expansion}
    \begin{split}
    \langle \hat W(t)^\dagger \hat V \hat W(t)\rangle & = \langle \hat V\rangle + i\phi \langle[\hat A(t),\hat V]\rangle \\
    & +\phi^2 \left( \langle \hat A(t) \hat V \hat A(t) \rangle - \frac{1}{2}\langle\{\hat A(t)^2,\hat V\}\rangle\right) \\
    & + \mathcal{O}(\phi^3)
\end{split}
\end{equation}
where the contribution quadratic in $\phi$ contains an OTOC of Hermitian operators: $\hat A$ acts at time $t$, then $\hat V$ acts at time $0$ followed by $\hat A$ acting at time $t$. This ordering is not compatible with the ordering of operators on a Schwinger-Keldysh contour where the ordering is from small to long and back to small times, hence the name OTOC. In general, it is not straight forward to separate the OTOC part from the additional terms that appear. However, in the case where the initial state $\ket{\psi_0}$ is an eigenstate of the operator $\hat V$ matters simplify considerably and the squared commutator can be obtained from measurements of the quadratic response after time reversal as we discuss in the following.

For $\hat V \ket{\psi_0} = \Lambda \ket{\psi_0}$ the squared commutator becomes (expectation values are always with respect to $\ket{\psi_0}$)
\begin{equation}
\begin{split}
    C(t)&=\langle [\hat A(t),\hat V]^\dagger [\hat A(t),\hat V]\rangle \\
    & = -2\Lambda\langle\hat A(t)\hat V\hat A(t)\rangle + \langle\hat A(t)\hat V^2\hat A(t)\rangle + \Lambda^2\langle\hat A^2(t)\rangle \\
    & = \left\langle\hat A(t) (\hat V-\Lambda )^2\hat A(t)\right\rangle \,.
\end{split}
\end{equation}
For the response of the n$th$ moment of $\hat V$ the commutator term [linear term in $\phi$ in Eq.~\eqref{eq:phi_expansion}] vanishes such that up to second order in $\phi$ one obtains
\begin{equation}
\begin{split}
    \langle \hat W(t)^\dagger \hat V^n \hat W(t)\rangle &= \Lambda^n + \phi^2 \left[\langle\hat A(t)\hat V^n\hat A(t)\rangle - \Lambda^n \langle\hat A^2(t)\rangle \right]   \\
    & \equiv \Lambda^n + \phi^2 \Gamma^{(2)}_{\hat V^n}
\end{split}
\end{equation}
where we have defined the quadratic response $\Gamma^{(2)}_{\hat O}$ of an operator $\hat O$. 
Thus the squared commutator can be obtained by measuring the quadratic response of the first and second moment of $\hat V$ in the time-reversal protocol described above as
\begin{equation}
    C(t) = -2\Lambda\Gamma^{(2)}_{\hat V} + \Gamma^{(2)}_{\hat V^2}
\end{equation}
In the case at hand one could for example initialize the system in the state where all particles are in the $m_F=0$ mode and measure $\hat V=\hat N_0$.

We now turn to a concrete experimental implementation of the model studied above. As mentioned in Sec.~\ref{sec:intro}, it can be straightforwardly realized using a BEC of $^{87}$Rb atoms in the $F=1$ hyperfine manifold in a tightly confining trap. The integrable case $r=0$ has been realized experimentally and its dynamics studied by a number of groups \cite{StamperKurn2013,Gerving2012,Hamley2012, Klempt2018,Oberthaler2018,Linnemann2016}. The integrability breaking term can be implemented by continuously applying an rf driving field where $r$ corresponds to the Rabi frequency induced by the drive. rf-driving is commonly used to implement spin rotations that are fast compared to the interacting dynamics in order to read out different components of the spin \cite{Kunkel2019}. By reducing the modulation amplitude of the magnetic field the Rabi frequency can be adjusted such that $r$ is widely tunable. The interaction term in the Hamiltonian results from s-wave collisions and is typically on the order of $gN/2\pi\approx-2\,$Hz. The parameter $q$ can be adjusted by microwave dressing to be of the same order \cite{Oberthaler2018}.

The challenge for experiments lies in measuring the OTOC, which involves time-reversing the unitary dynamics of the system as illustrated by the protocols discussed above. Indeed, for $^{87}$Rb the nature of the collisional interactions between the different hyperfine components allows one to do exactly this. In the $F=2$ hyperfine manifold the sign of $g$ in the spin-changing collision term is opposite to $F=1$ and significantly larger. The other parameters of the Hamiltonian can be adjusted freely through external control fields. Thus, reversal of the dynamics can be accomplished by transferring the BEC into the $F=2$ manifold and running the dynamics under the sign-reversed Hamiltonian for a time that has to be adjusted in order to compensate for the different magnitude of the Hamiltonian parameters. The additional two levels ($m_F=\pm 2$) in the $F=2$ manifold can be tuned out of resonance by adjusting the magnetic field and thus the quadratic Zeeman shift, or by off-resonant microwave dressing, resulting in effective three-mode dynamics.

Finally, we discuss a number of recent proposals for measuring OTOCs. We tried to adapt them to the proposed implementation, however, all of them either measure OTOCs with respect to thermal (or even infinite temperature) states or have technical requirements that seem extremely challenging for spinor BECs. Protocols using randomized measurements \cite{Vermersch2018} only allow to determine OTOCs of the maximally mixed state. The scheme proposed in Ref.~\cite{Bohrdt2017} requires preparing two identical copies of the system and detecting the atom number parity, while \cite{Yao2016} additionally requires an ancillary systems that controls the sign of the Hamiltonian and similarly for \cite{Guanyu2016} which is designed to detect OTOCs of thermal states.

\section{Conclusions and outlook} \label{sec:conclusion}

This study shows that a three-mode bosonic system with spin-changing collisions and tunable integrability breaking rf-coupling term can serve as a showcase system for studying classical and quantum chaos. In the classical limit the system transitions from regular to chaotic phase space dynamics as the strength of the rf-coupling is increased from zero to $\sim 1$ in units of the characteristic energy scale of the interactions. This transition is also confirmed by the Lyapunov exponents which are positive in the emerging chaotic regions of phase space. The statistics of eigenenergies, a traditional indicator of quantum chaos, behaves as expected and changes from Poissonian to Wigner-Dyson as the integrability breaking term is tuned. For OTOCs, which have been proposed as faithful indicators for quantum chaos in the semi-classical regime, we find faster growth for initial states close to the separatrix where the classical dynamics becomes chaotic even for small integrability breaking.
The characteristic exponential growth that is expected due to the connection between the OTOC and the Lyapunov exponent in the classical limit is visible in an appreciable time window for large particle numbers ($\gtrsim 10^3$). 
A direct comparison of the resulting growth exponent to the largest Lyapunov exponent defined in the standard way is not meaningful here, which can be understood by considering the semi-classical truncated Wigner expression \eqref{eq:OTOC_TWA} for the OTOC. This form shows that an OTOC of specific operators will depend on specific elements of the stability matrix, or fundamental matrix, of the classical dynamics, which are in general not expected to grow exponentially with the largest Lyapunov exponent.

The considered model can be implemented experimentally, potentially giving access to OTOCs beyond the Heisenberg time for much larger particle numbers than what is reachable by exact diagonalization. Time-reversal may be achieved in BECs of $^{87}$Rb atoms by exploiting the collisional properties in different hyperfine states. This enables the measurement of OTOCs with respect to initial states that are eigenstates of the measured operator. Specifically the squared commutator, which has a particularly direct semi-classical meaning, can be extracted for initial coherent states which are easy to prepare experimentally. The protocol for extracting the squared commutator requires measuring the quadratic response with respect to a perturbing generalized rotation. More straightforward protocols for testing the reversibility and sensitivity of the many-body dynamics with respect to perturbations also allow measuring quantities that are of interest for semi-classical physics and quantum chaos \cite{Schmitt2019}. Therefore, in the future, we will seek to clarify the meaning of the resulting more general types of OTOCs in the context of operator spreading and thermalization. Intuitively, these objects should still be indicators of ergodicity and emerging operator complexity, and are in addition related to quantum Fisher information \cite{Gaerttner2018}, enabling connections to entanglement. 

\acknowledgements
We thank Thomas Gasenzer, Benjamin Geiger, Quirin Hummel, Daniel Linnemann, and Markus Oberthaler for discussions. Supported by the DFG Collaborative Research Center SFB1225 (ISOQUANT).

\appendix 

\section{Truncated Wigner Approximation} \label{app_sec:TWA}

\begin{figure} 
	\includegraphics{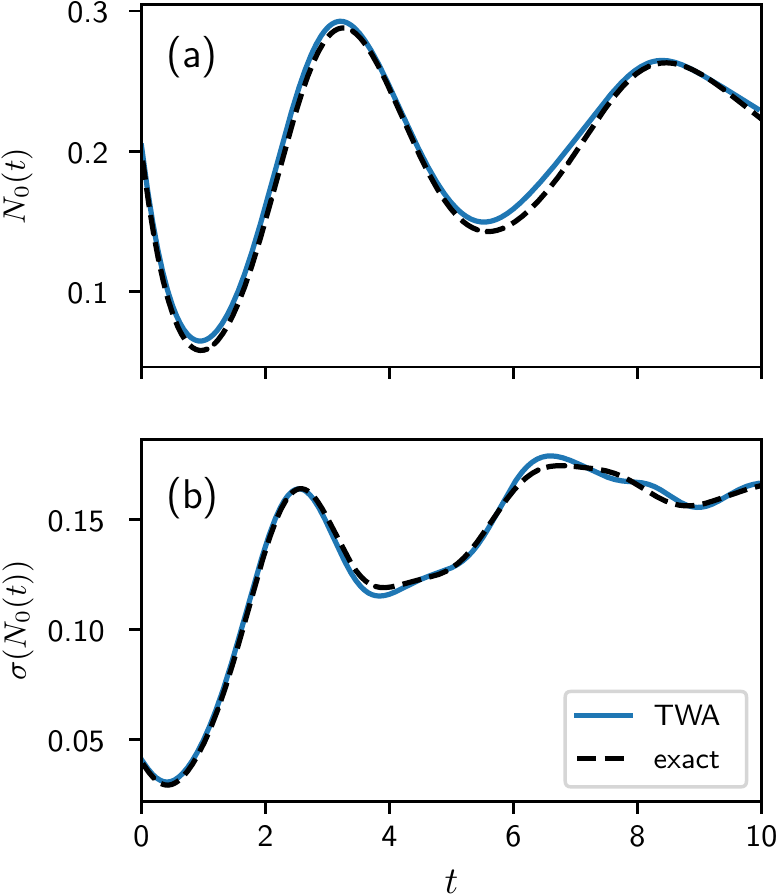} 
	\caption{Comparison of the time evolution of the expectation value of $\hat{N}_0$ (a) and its standard deviation (b) calculated with TWA (solid blue line) to the exact quantum mechanical calculation (dashed black line) for $N = 100$ particles. As initial state we used the classically chaotic state from Fig.~\ref{fig:CoherentStates} along with the same parameters. For the TWA we averaged over 1000 samples. We also compared the TWA to the exact quantum calculations for different observables and initial states yielding similar results, for classically regular dynamics the agreement is even better.}
	\label{fig:TWAvsExact}
\end{figure}

The TWA relies on a phase space representation of quantum states where expectation values of quantum operators can be calculated by evaluating their corresponding Weyl symbols averaged over phase space weighted with the Wigner function where the latter is the Weyl symbol of the density matrix of the initial state. As we are interested in calculating OTOCs, we first note that the Weyl symbol of the commutator $\left[ \hat{A}, \hat{B} \right]$ is given by 
\begin{equation}
\begin{split}
    \left[ \hat{A}, \hat{B} \right]_W &= i \hbar\ind{eff} \left\lbrace A_W, B_W \right\rbrace_{MB} \\
    &=  i \hbar\ind{eff} \left\lbrace A_W, B_W \right\rbrace + \mathcal{O}\left(\hbar\ind{eff}^2\right)  
\end{split}
\end{equation}
where $\left\lbrace \dots \right\rbrace_{MB}$ denotes the Moyal bracket,  $\left\lbrace \dots \right\rbrace$ the Poisson bracket and $O_W$ is the Weyl symbol of the quantum operator $\hat{O}$ \cite{Polkovnikov2010}. $\hbar\ind{eff}$ quantifies the minimal phase space volume that a state can be confined to due to Heisenberg's uncertainty relation, which corresponds to the phase space volume of a coherent state, $\hbar\ind{eff}\sim 1/N$ in our case.
Working directly in the $\zeta$ representation (see beginning of section \ref{sec:model}) where the real and imaginary parts of $\zeta_i$ act as pairs of conjugate variables, the Poisson bracket is defined as 
\begin{equation}
    \left\lbrace A, B \right\rbrace \equiv \sum_{i = 1, 0, -1} \frac{\partial A}{\partial \zeta_i^R} \frac{\partial B}{\partial \zeta_i^I} - \frac{\partial A}{\partial \zeta_i^I} \frac{\partial B}{\partial \zeta_i^R}
\end{equation}
i.e. we have for $A = \rho_0(t) = \left(\zeta_0^R(t)\right)^2 +  \left(\zeta_0^I(t)\right)^2$ and $B = \rho_0(0)$ up to order $\hbar\ind{eff}^2$
\begin{align}
    \begin{split}
    &\left\langle\left| \left[ \hat{\rho_0}(t), \hat{\rho_0}(0) \right] \right|^2\right\rangle = \hbar\ind{eff}^2 \left\langle \left|\left\lbrace \rho_0(t), \rho_0(0) \right\rbrace \right|^2 \right\rangle_W = \\
    &\hbar\ind{eff}^2 \left\langle \left| \frac{\partial \rho_0(t)}{\partial \zeta_0^I} \frac{\partial \rho_0(0)}{\partial \zeta_0^R} \right|^2 \right\rangle_W = 
    \hbar\ind{eff}^2 \left\langle \left| 2 \zeta_0^R(0) \frac{\partial \rho_0(t)}{\partial \zeta_0^I} \right|^2 \right\rangle_W
    \end{split}
\end{align}
where $\langle \dots \rangle_W$ denotes the phase space average weighted by the Wigner function. In the second step we used that we can always choose $\zeta_0$ to be real initially by choosing an appropriate global phase. The derivative ${\partial \rho_0(t)}/{\partial \zeta_0^I}$ is evaluated numerically by evolving two trajectories initially separated by ideally infinitely small $d_0$ in the imaginary part of the $\zeta_0$ component and evaluating the difference in $\rho_0$ at each time step numerically integrating the mean field e.o.m. \eqref{eq:MFeom}. Alternatively, we can rewrite 
\begin{equation}
    \frac{\partial \rho_0(t)}{\partial \zeta_0^I} = 2 \left\lbrace \zeta_0^R(t) \frac{\partial \zeta_0^R (t)}{\partial \zeta_0^I} + \zeta_0^I(t) \frac{\zeta_0^I(t)}{\partial \zeta_0^I} \right\rbrace
\end{equation}
where now the derivatives can be evaluated as elements of the fundamental matrix as described in Sec.~\ref{subs:LyapunovExp}. We again checked that the two methods yield the same results for small enough initial deviations $d_0$ in the case of the two trajectory calculation (or equivalently, for short enough times).
The Wigner function of the initial coherent state is well approximated by a Gaussian with mean and variance matching the expectation value and variance of the initial state w.r.t. the quantum operators. Technically, we sample from $\ket{(x_1+ix_2)/2,N,(x_3+ix_4)/2}$ where all $x_i$ are independently drawn from a Gaussian distribution with zero mean and standard deviation $1/\sqrt{N}$ to approximate the Wigner function  of the initial state $\ket{0,N,0}$. To get the Wigner function of an arbitrary initial coherent state, we rotate the sampled states using spin operators as generators of the rotation. 

We checked that the sampling of arbitrary initial states as described above works by comparing the expectation values and variances of different spin operators calculated within TWA to the exact quantum mechanical calculations for up to $N = 100$ particles, see Fig.~\ref{fig:TWAvsExact}. We also checked that in the case of the two trajectory calculation $d_0$ is small enough s.t. the difference quotient is independent of $d_0$ and matches the quantum calculations after appropriately scaling the amplitude (see inset of Fig.~\ref{fig:OTOC}). We can confirm that the TWA results match the exact quantum calculations up to $t \approx 2.5$ for $N = 100$ particles and expect that the approximation holds up to longer times with increasing $N$ as $t\sim \log N$.

%

\end{document}